Dirac dispersion generates large Nernst effect in Weyl semimetals

Sarah J. Watzman[1], Timothy M. McCormick[2], Chandra Shekhar[3], Shu-Chun Wu[3], Yan Sun[3], Arati Prakash[2], Claudia Felser[3], Nandini Trivedi[2], Joseph P. Heremans[1,2,4]*

[1]Department of Mechanical and Aerospace Engineering, The Ohio State University, Columbus, OH, USA, 43210.
[2]Department of Physics, The Ohio State University, Columbus, OH, USA, 43210.
[3]Max Planck Institute for Chemical Physics of Solids, Dresden, Germany, 01187.
[4]Department of Materials Science and Engineering, The Ohio State University, Columbus, OH, USA, 43210.

**Abstract**

Weyl semimetals expand research on topologically protected transport by adding bulk Berry monopoles with linearly dispersing electronic states and topologically robust, gapless surface Fermi arcs terminating on bulk node projections. Here, we show how the Nernst effect, combining entropy with charge transport, gives a unique signature for the presence of Dirac bands. The Nernst thermopower of NbP (maximum of 800 $\mu VK^{-1}$ at 9 T, 109 K) exceeds its conventional thermopower by a hundredfold and is significantly larger than the thermopower of traditional thermoelectric materials. The Nernst effect has a pronounced maximum near $T_M=90\pm20$ K$=\mu_0/k_B$ ($\mu_0$ is chemical potential at T=0 K). A self-consistent theory without adjustable parameters shows that this results from electrochemical potential pinning to the Weyl point energy at $T \geq T_M$, driven by charge neutrality and Dirac band symmetry. Temperature and field dependences of the Nernst effect, an even function of the charge polarity, result from the intrinsically bipolar nature of the Weyl fermions. Through this study, we offer an understanding of the temperature dependence of the position of the electrochemical potential vis-à-vis the Weyl point, and we show a direct connection between topology and the Nernst effect, a potentially robust experimental tool for investigating topological states and the chiral anomaly.

**Introduction**

Weyl fermions, which are chiral, massless solutions to the Dirac equation, were predicted to exist in wide-ranging candidate materials called Weyl semimetals (WSM) (*1-5*). Angle-resolved photoemission spectroscopy (ARPES) and scanning tunneling spectroscopy confirmed their existence experimentally in transition metal monopnictides (*6-13*) and dichalcogenides (*14,15*). Weyl points are magnetic monopoles in momentum space that come in opposite-chirality pairs. Weyl semimetals must break either time-reversal or inversion symmetry to separate the two sets of linearly dispersing bands with opposite Berry curvature in momentum space. An unusual WSM characteristic is the presence of Fermi arcs in surface Brillouin zones that terminate at Berry monopole projections.

The Weyl nodes' Berry curvature leads to a variety of electronic transport phenomena in WSMs including negative longitudinal magnetoresistance (*16*), a huge magnetoresistance, and ultrahigh mobility (*17*). Preliminary theoretical studies predict that bulk Weyl nodes (*18,19*) and surface Fermi arcs (*20*) have signatures in magnetothermal and thermomagnetic transport, beyond that of classical electrical transport in semimetals, where electrons and holes co-exist, as seen with nanostructuring in NbP (*21*). The Seebeck and Hall effects measure the difference between charges carried by both types of carriers, whereas entropy transport and the Nernst effect measure their sum. Additionally, the anomalous Nernst coefficient measures the Berry curvature directly on the Fermi surface (*22*), contrasting with the Hall conductivity, which averages over the Fermi surface volume (not applicable to NbP, which has no net Berry curvature) (*23*). Experimental thermomagnetic data in Dirac semimetals exists (*23-26*), including the magnetic-field dependent adiabatic Seebeck coefficient of NbP (*27*), but quantitative explanations for the data are lacking. The WSM Nernst effect remains unexplored theoretically and experimentally. Here, we show that the symmetry between the electron-like and hole-like Dirac band portions results in *(i)* a characteristic temperature- and magnetic-field-dependent Nernst thermopower, and *(ii)* a relation between the Nernst and Seebeck coefficients that are unique Dirac band signatures.

We report and explain the single-crystal NbP isothermal Nernst and Seebeck effects with data on samples with a large, unsaturated magnetoresistance and ultrahigh mobilities. The NbP band structure consists of 24 Weyl points and several trivial pockets (*17*). In the low-temperature limit, the electron or hole charges residing in the trivial bands and/or in chemical defects in the samples, including vacancies or unintentional aliovalent impurities, determine the NbP chemical potential, $\mu_0$. The observation that, with increasing temperatures, the electrons/holes in the symmetric Dirac bands dominate transport as the electrochemical potential $\mu(T)$ moves toward the Dirac point is a key finding. The space-charge neutrality condition dictates the electrochemical potential pinning at the Dirac point when the thermally excited carrier density exceeds that of residual carriers. The second finding in this manuscript is that the Nernst effect is particularly sensitive to this.

The Nernst effect is a decreasing function of temperature in classical semimetals like Bi (*28*), except in the phonon-drag regime, which is excluded here *(27)*. But, it becomes a non-monotonic function of temperature with a maximum at $T_M=\mu_0/k_B$ when the bands have a Dirac-like dispersion and are symmetric around the Dirac point. The Nernst thermopower of charge carriers in Dirac bands exceeds 800 µV/K at 9 T, 109 K, and does not saturate within our measurement range. This Nernst thermopower is two orders of magnitude larger than the Seebeck coefficient: the isothermal Seebeck coefficient is $|\alpha|\leq 8$ µV/K under the same circumstances; the adiabatic Seebeck coefficient (*27*) becomes much larger in the presence of a transvere magnetic field because it contains a large contribution from the Nernst effect. Indeed, the Seebeck coefficients arising in the upper and lower parts of the Dirac band are each ascribed a partial Seebeck coefficient that is an odd function of the charge-carrier polarity; thus, the total Seebeck coefficient averages two counteracting contributions. In contrast, the Nernst coefficient is an even function of the charge-carrier polarity, entering once as in the Seebeck coefficient and again via the Lorentz force. Therefore, in the total Nernst coefficient, the opposing-polarity charge-carrier contributions add, and the Seebeck effect is smaller than the Nernst effect even in the presence of an applied magnetic field. We show further that the non-monotonic temperature dependence is specific to Dirac bands: for $T<T_M$, the Nernst coefficient will increase with increasing temperature as $\mu(T)$ approaches 0, then decrease, as in Bi (*28*) at $T>T_M$. We provide experimental evidence for this and present a quantitative theory to explain the data.

**Results and Discussion**

*Experimental isothermal thermomagnetic transport in NbP*

NbP single crystals are used from the same source as References (*17*) and (*27*). The single crystals were characterized using resistivity, Hall effect, thermal conductivity, specific heat, de Haas–van Alphen oscillations, and Shubnikov–de Haas oscillations. The results of these characterizations are found in the Supplementary Material. With a magnetic field applied along the z-axis <001>, the isothermal Nernst thermopower $\alpha_{xyz}$ is defined as the transverse electric field measured along the crystal's y-axis, with a temperature gradient and heat flux applied along the x-axis. The isothermal Seebeck coefficient $\alpha_{xxz}$ measures the longitudinal electric field along the x-axis under the same circumstances:

$$\alpha_{xyz} \equiv \left. E_y(H_z) \middle/ -\nabla_x T \right|_{\nabla_y T=0}$$
$$\alpha_{xxz} \equiv \left. E_x(H_z) \middle/ -\nabla_x T \right|_{\nabla_y T=0}$$
(1)

The Nernst coefficient is defined as $N_{xyz} \equiv \left. dα_{xyz}/dH_z \right|_{\nabla_y T=0, H_z}$. These quantities are measured on a sample mounted to set $\nabla_y T = 0$, which is the definition of isothermal (as opposed to adiabatic) measurements (*29*). In the present experiment, the isothermal sample mount is described in Materials and Methods. The distinction between isothermal and adiabatic measurement conditions affects the boundary conditions of the sample and thus the derivation of the Nernst thermopower and the Seebeck coefficient in magnetic field *(29)*. The thermomagnetic tensor relates the electric field values measured along the *x* and *y* directions to the applied temperature gradients as:

$$\begin{pmatrix} E_x \\ E_y \end{pmatrix} = \begin{pmatrix} α_{xxz} & α_{yxz} \\ α_{xyz} & α_{yyz} \end{pmatrix} \begin{pmatrix} -\nabla_x T \\ -\nabla_y T \end{pmatrix} \tag{2}$$

where the $α$'s are as defined in Eq. 1. The more commonly used mount is the adiabatic mount *(27)*, in which the sides of the sample along *z* are thermally isolated. In samples in which the electronic contribution to the thermal conductivity is appreciable and is subject to a Hall effect, as is the case here, the measurement of the adiabatic Nernst effect measures $E_y / -\nabla_x T = α_{xyz} + α_{yyz} \nabla_y T / \nabla_x T$, and is thus a mixture of $α_{xyz}$ and $α_{yyz}$, the longitudinal thermopower along *y*. In the adiabatic mount, the magnetic-field dependent longitudinal thermopower $E_x / -\nabla_x T = α_{xxz} + α_{yxz} \nabla_y T / \nabla_x T$ contains a contribution of the Nernst effect, which in NbP is much larger than the Seebeck effect. All Nernst effect theories consider only the isothermal Nernst ($α_{xyz}$) and Seebeck ($α_{xxz}$) effect as defined by Eq. 1.

The experimental results of the Nernst effect measurements are shown in Figs. 1A (magnetic-field dependence of the Nernst thermopower $α_{xyz}$ at discrete temperatures) and 1B (temperature dependence of the Nernst coefficient $N_{xyz}$). $α_{xyz}$ is an odd function of $H_z$ with a higher slope near zero magnetic field than at high field. We see an unsaturated large Nernst thermopower, with a maximum exceeding 800 μV/K at 9 T, 109 K, which is 2-4 times larger than the maximum thermopower of conventional, commercial thermoelectric semiconductors. The Nernst coefficient $N_{xyz}$ temperature dependence taken at low field ($H_z$<2 T) and high field (magnitude between 3 T and 9 T) is non-monotonic, with a maximum around $T_M$~50 K for low-field $N_{xyz}$ and $T_M$~90 K for high-field $N_{xyz}$. Fig. 2 shows the temperature dependence of $α_{xxz}$, which is nearly two orders of magnitude smaller than the Nernst thermopower $α_{xyz}$(9 T) near 100 K, and its absolute value is much smaller than the high-field Nernst effect at all temperatures.

The $\alpha_{xxz}$ temperature dependence has a broad minimum around $T_M$ ranging from 60 to 100 K. No magnetic-field dependence is observed for $\alpha_{xxz}(H_z)$ within instrument sensitivity, despite repeated attempts. This result does not contradict prior measurements (27) of the adiabatic magneto-Seebeck coefficient $\alpha_{xxz} \equiv \left. \frac{E_y(H_z)}{-\nabla_x T} \right|_{\nabla_y T \neq 0}$ because those reflect a contribution from $\alpha_{xyz}$ in the data. The properties of the present sample's Fermi surface exclude phonon drag as a possible source for observed non-monotonicity in $\alpha_{xxz}$ and $\alpha_{xyz}$ at temperatures exceeding 14 K (27).

At low temperatures (inset of Fig. 1A), the Nernst thermopower exhibits Shubnikov–de Haas oscillations. The periods observed in Shubnikov-de Haas and de Haas-van Alphen oscillations observed in the magnetization (see Supplemental Materials) correspond very well, within the error bars of our measurements, even though they were obtained on different samples that may have had different defect densities and thus different residual doping levels. These results are also similar to those found by J. Klotz et al. (30). The de Haas-van Alphen oscillation periods for $H$ parallel to all three axes and the corresponding values for the area of the Fermi surfaces and values for $\mathbf{k}_F$ calculated assuming that the cross-sections are circles are reported in Table 1. These periods are used in conjunction with the DFT calculations (further explanation in Materials and Methods) to derive the position of the electrochemical potential vis-à-vis the Weyl points in the zero-Kelvin limit, which is determined to be $\mu_0 \approx$ -8.2±2 meV below the main Weyl bands' Dirac point. This sets the Fermi level in the zero-K limit in the valence band, which is consistent with the positive thermopower observed in Fig. 2 at $T < 25$ K.

*Obtaining a theoretical model for inversion-symmetry-breaking Weyl semimetals*

The theory is based on a WSM Hamiltonian (31) given by

$$\hat{H}^{IB}(\mathbf{k}) = -(m(1-\cos^2(k_y a) - \cos(k_z a)) + 2t_x(\cos(k_x a) - \cos(k_0 a)))\hat{\sigma}_1 - 2t\cos(k_y a)\hat{\sigma}_2 - 2t\sin(k_z a)\hat{\sigma}_3$$

(3),

which supports four Weyl nodes located at $\mathbf{k} = (\pm k_0, \pm\pi/2, 0)$. The parameters $t$, $t_z$, and $m$ set the bandwidth of the bands away from the Weyl points, and $a$ is the lattice spacing. Fig. 3A plots the energy band structure for a chosen set of $t$, $t_z$, and $m$. Fig. 3B gives a two-dimensional picture of one Dirac band. The model is solved self-consistently for the chemical potential $\mu(T)$, which moves to the nodal energy with increasing temperature on a scale of $T/{\mu_0/k_B}$ as illustrated in Fig. 3C. This is crucial to the temperature-dependent Nernst effect.

The calculation of the thermomagnetic tensor elements in the Boltzmann formalism (*18,19*) yields the following components:

$$L_{xx}^{ET} = e \int \frac{d^3k}{(2\pi)^3} v_x^2 \tau \frac{E-\mu(T)}{T}\left(-\frac{\partial f_0}{\partial E}\right)(c_x - D) \quad (4)$$

$$L_{xy}^{ET} = -e \int \frac{d^3k}{(2\pi)^3} (v_y^2 c_y + (c_y - D)v_x v_y)\tau \frac{E-\mu(T)}{T}\left(-\frac{\partial f_0}{\partial E}\right) - \frac{k_B e}{\hbar}\int \frac{d^3k}{(2\pi)^3}\Omega_z s_\mathbf{k} \quad (5)$$

$$L_{xx}^{EE} = e^2 \int \frac{d^3k}{(2\pi)^3} v_x^2 \tau \left(-\frac{\partial f_0}{dE}\right)(c_x - D) \quad (6)$$

and

$$L_{xy}^{EE} = -e^2 \int \frac{d^3k}{(2\pi)^3}\left(v_y^2 c_y + (c_y - D)v_x v_y\right)\tau\left(-\frac{\partial f_0}{\partial E}\right) - \frac{e^2}{\hbar}\int \frac{d^3k}{(2\pi)^3}\Omega_z f_0 \quad (7).$$

Here $c_x$ and $c_y$ are corrective factors (*19*) from the magnetic field and Berry curvature, and $D(\mathbf{B},\mathbf{\Omega}) = \left(1 + \frac{e}{\hbar}\mathbf{B}\bullet\mathbf{\Omega}\right)^{-1}$ originates from the modification of the phase-space volume element in a magnetic field. Because the Hamiltonian in Eq. 3 preserves time-reversal symmetry, the second integrals in Eqs. 5 and 7 vanish.

Using Eqs. 4-7, the Nernst thermopower can be calculated (see Materials and Methods) and is given by

$$\alpha_{xyz} = \frac{E_y}{-\nabla_x T} = \frac{L_{xx}^{EE} L_{xy}^{ET} - L_{xy}^{EE} L_{xx}^{ET}}{\left(L_{xx}^{EE}\right)^2 + \left(L_{xy}^{EE}\right)^2} \quad (8).$$

Results are shown in Fig. 1C as a function of magnetic field and in Fig. 1D as a function of temperature. The magnetic field dependence shows several salient features, notably a change in monotonicity for different temperatures and distinct slopes at high and low field. The origin of the non-monotonicity at low temperatures arises from the second term in the numerator of Eq. 8. At low temperatures, the conventional Hall response $L_{xy}^{EE}$ dominates the Nernst response. At higher temperatures, the Nernst response is given by

$$\alpha_{xyz} \approx \frac{L_{xy}^{ET}}{L_{xx}^{EE}} \quad (9).$$

Two distinct regions are present with different slopes at low field and at high field. For a reasonable choice of lattice spacing $a$, the field scale delineating the two regions matches that found in our experimental results on NbP.

The linear dispersion of the Dirac bands results in a quadratic dependence of the density of states' energy dependence. If the relaxation time also has an energy dependence that is a power law of energy, $\tau(\varepsilon) = \tau_0 \varepsilon^\lambda$ with $\lambda$ an integer, the Fermi functions that enter the integrals of Eq. 4-7 at zero magnetic field have analytical solutions, obviating the need for the Bethe-Sommerfeld expansion (*32*) valid only in the degenerate limit. Therefore, analytic solutions can be derived for the thermopower and the low-field Nernst coefficient of symmetric Dirac bands that are valid at all temperatures; for the case of the energy-independent relaxation time, they are:

$$|\alpha_{xxz}| = \frac{2\pi^2}{3} \frac{k_B}{e} \frac{\mu}{k_B T} \left[ \frac{\left(\frac{k_B T}{\mu}\right)^2}{1 + \frac{\pi^2}{3}\left(\frac{k_B T}{\mu}\right)^2} \right]$$

$$\lim_{H_z \to 0} N_{xyz} = \frac{\pi^2}{3} \frac{\tau v_F^2}{cT} \left[ \frac{\left(\frac{k_B T}{\mu}\right)^2}{1 + \frac{\pi^2}{3}\left(\frac{k_B T}{\mu}\right)^2} \right]$$

(10).

These functions describe quantitatively most of the important features of the low-field Nernst and Seebeck coefficients reported in Figs. 1 and 2, including the observed maxima at $T_M \simeq \frac{\sqrt{3}}{\pi} \frac{\mu_0}{k_B}$, above which $N_{xyz}$ decreases following a $1/T$ law. These functions are plotted in the Supplementary Materials.

*Comparison of experimental data and theoretical model for Nernst effect in NbP*

The full model of Eq. 8, including the magnetic field and Berry curvature effects, without any adjustable parameters, is compared to the experimental data. The magnetic field scales with the magnetic length and the lattice spacing. The temperature scale is set by $\mu_0$. The scattering time $\tau$ is assumed to be energy- and temperature-independent; appropriate values of $v_F \sim 2 \times 10^5$ m s$^{-1}$ (determined using values from density functional theory and ARPES) and resistivity (see Supplementary Materials) yield a value of $\tau \sim 10^{-13}$ s. The calculation results are shown in Fig. 1C for $\alpha_{xyz}$ magnetic-field dependence at discrete temperatures and Fig. 1D for $N_{xyz}$ temperature dependence. The calculated field and temperature dependences show remarkable agreement to experimental results, apart from the sign change that theory predicts for low-field $N_{xyz}$ below 40 K. The amplitudes agree within a factor of 4; this is remarkable given the uncertainty of $\tau$. Fig. 1 shows the high-field Nernst effect peaks at $T_M \simeq \frac{\mu_0}{k_B}$ in both theory and experiment, whereas near zero-field $N_{xyz}$ peaks around $T_M \simeq \frac{\sqrt{3}}{\pi} \frac{\mu_0}{k_B}$ in both theory and experiment. The Seebeck coefficient also is calculated to have an absolute value maximum at $T_M \simeq \frac{\sqrt{3}}{\pi} \frac{\mu_0}{k_B}$. The $\alpha_{xxz}$ and $\alpha_{xyz}$ extrema originate from the chemical potential temperature dependence. As the chemical potential shifts to the nodal energy with increasing temperature, the effects of the Berry curvature strengthen, maximizing the Nernst effect, shown via experiment and theory. At temperatures above $T_M$, the Fermi function derivative broadens, leading states away from the Weyl nodes to contribute more to transport, lowering the Nernst thermopower and Nernst coefficient.

*Consideration of trivial pockets*

Trivial pockets in the Fermi surface have been ignored in the model above, whereas they are known to exist from band structure calculations valid in the zero-Kelvin limit. When their contributions are included in the calculations of the Nernst and Seebeck effects (see Supplemental Materials), the model no longer follows the experiments. However, in addition to the agreement between experiment and simple theory that disregards them, other observations indicate that they do not contribute to transport above 100 K. First, the high-temperature behavior of the conventional thermopower (Fig. 2) tends to zero, the value it reaches rigorously in a Dirac band when $\mu=0$ (Eq. (10)). If trivial pockets were to contribute to the thermopower, as is the case for Bi *(28)*, $\alpha(T)$ would become metallic and increase monotonically with increasing temperature. Second, the near-perfect compensation between electron and hole pockets also explains the galvanomagnetic properties *(17)*. The band structure temperature dependence may be responsible for this: band gaps of most semiconductors increase (PbTe) or decrease (GaAs) by ~100 meV from cryogenic to room temperatures. Given that the energy scale for the Dirac fermions relevant to transport is an order of magnitude less, this is possibly the origin of the disappearing importance of trivial pockets with increasing temperature.

*Implications for applications in thermoelectric energy generation*

The maximum Nernst thermopower found in this study, $\alpha_{xyz}$(9 T, 109 K) ~ 800 µV/K, is a remarkable result, surpassing the conventional thermopower of traditional thermoelectric semiconductors (~200–300 µV/K). Although the resistivity of NbP at 9 T, 100 K, is 1.5 x $10^{-5}$ Ω m because of the large magnetoresistance, the geometry of $\alpha_{xyz}$ is favorable for thermoelectric technology. The output of a transverse Nernst thermoelectric generator or Ettingshausen cooler scales intrinsically with device size *(33,34)*, does not require separate n- and p-type legs, and can function over large temperature differences without staging or cascading, offering advantages in this geometry.

**Summary**

In conclusion, we experimentally and theoretically explored thermomagnetic transport in the inversion-symmetry-breaking WSM NbP. Using isothermal measurements, we experimentally characterized the NbP Nernst and Seebeck effects. Specifically, two regimes in Nernst thermopower are seen: a high-field Nernst for $H_z>|3\text{ T}|$, and a low-field Nernst for $H_z<|2\text{ T}|$; both have non-monotonic temperature dependences. The theory that explains these properties quantitatively shows them to be signatures of transport in Dirac bands. The Nernst thermopower $\alpha_{xyz}$(9T, 109 K) exceeds 800 µV/K, surpassing the Seebeck thermopower $\alpha_{xxz}$ by two orders of magnitude and also surpassing the thermopower of traditional thermoelectric semiconductors while utilizing a more favorable transverse geometry. Through this study, we offer an understanding of the temperature dependence of the position of the electrochemical potential vis-à-vis the Weyl point and show a direct connection between the Nernst effect and topology, a potentially robust mechanism for investigating topological states and the chiral anomaly.

**Methods**

*Isothermal thermomagnetic transport measurements*

A single-crystal sample of NbP, dimensions 2.25 x 1.29 x 0.52 mm, made using methods described in Ref. *(17)*, was characterized in this work. The sample was mounted as shown in Fig. 4 on a silicon backing plate to which a temperature gradient along *x* is applied. The silicon plate that underlies the sample's whole length and width is designed to short-circuit the sample thermally so that the temperature gradient established in the silicon also is imposed on the NbP, with $\nabla_y T \approx 0$, i.e. an isothermal mount *(29)*. This structure in turn is mounted in the Thermal Transport Option (TTO) on a Quantum Design Physical Property Measurement System (PPMS) modified for isothermal Nernst and Seebeck measurements. This mount differs fundamentally from the conventional adiabatic mount recommended by the manufacturer and used in Ref. *(27)*.

The magnetic field was applied parallel to the thin direction (labeled *z*), corresponding to the <001> crystal axis. The heat flux was applied along the *x*-axis, while the measurements of electric fields were either along the *x*-axis for the Seebeck coefficient $\alpha_{xxz}$ or along the *y*-axis for the Nernst thermopower $\alpha_{xyz}$. The (*x,y,z*) coordinates are orthonormal, but the orientations of *x* and *y* in the <001> plane are not identified. The sample was attached to the silicon base plate using GE varnish. Two insulated copper wires were attached on opposite ends of the sample using silver epoxy. Three 120 Ω resistive heaters were attached in series to a copper foil heat spreader, which was then attached to the base plate and sample using silver epoxy. A copper foil heat sink was used with a gold-plated copper plate for attachment to the puck clamp. Due to the small size of the sample, the thermometers themselves could not be attached to the sample. Gold-plated copper leads for temperature measurements were attached to the bottom of the silicon plate using silver epoxy at the location of the edges of the sample. The silicon base plate and sample are assumed to be isothermal at these points due to the high thermal conductivity of silicon and the fact that the temperature gradient is applied to the silicon itself. Gold-plated copper assemblies purchased from Quantum Design containing calibrated Cernox™ thermometers and voltage measurement wires were attached to the leads; the voltage wires were detached from these assemblies and soldered directly to the copper voltage wires on the sample. Measurements were taken at discrete temperatures between 2.5 K and 400 K, and magnetic fields were swept in both directions to a maximum magnitude of 9 T. Controls software was programmed using LabVIEW.

Measurements of the adiabatic Nernst thermopower are reported in the Supplementary Materials. They show discrepancies between isothermal and adiabatic Nernst coefficients on the order of 10% at 100 K, indicating the correspondence between adiabatic and isothermal Nernst thermopower data is quite within the experimental error. However, because the Nernst thermopower is much larger than the Seebeck thermopower, the adiabatic mount gives a very large magnetic field dependence to the latter, which is absent in the isothermal mount.

*Deriving the low-temperature electrochemical potential from de Haas-van Alphen oscillations*

The frequencies of the quantum oscillations observed in the sample magnetization (magnetization data shown in the Supplementary Materials) with $H \,//\, <001>$ and with $H \perp <001>$ are tabulated in Table 1. They are used to determine the position of the electrochemical potential at 5 K, where the data were taken, based on the density functional theory (DFT) calculations.

DFT calculations were performed by using the Vienna *ab-initio* Simulation Package (*35*) with a generalized gradient approximation (*36*). The tight binding model Hamiltonian was calculated by projecting Bloch states onto maximally localized Wannier functions (*37*), which was used for generating Fermi surfaces. In the zero-Kelvin limit, there are two types of Weyl points in NbP, one located in the $k_z=0$ plane and the other located around the $k_z= \pi/c$ plane along the <001>; the latter is the type of Weyl point contributing to transport and is illustrated in Fig. 5 (*38,39*). We calculated the quantum oscillation periods for both the electron and hole pockets with the magnetic field applied both along and perpendicularly to the <001> direction, as a function of the hypothetical position of the Fermi level (the electrochemical potential at 0 K), $\mu_0$. The results are shown in Fig. 6 for the magnetic field aligned with either the <001> axis or perpendicularly to this axis. The experimentally observed periods of the oscillations are then projected as horizontal lines in Fig. 6. The Fermi level is located a few meV from the Weyl points, while the accuracy of the DFT calculations cannot be expected to be of that order. Therefore, it is not realistic to expect a perfect fit of all the experimental frequencies of oscillations with all the calculations. Fig. 6 shows, as a grey hatched area, that a range of value of $\mu_0 = $ -8.2±0.7 meV satisfies most of the observed oscillations very well. This then is the value that is used in the main text to adjust the scales of calculated and experimental results in Fig. 1 without adjustable parameters.

*Obtaining temperature dependence of the analytic thermopower and low-field Nernst coefficient*

The continuum form of the thermomagnetic transport tensors is from R. Lundgren et al. (*40*), where $\omega_c \equiv eBv_F^2/cE$, $\Omega$ is the Berry curvature, $f_0$ is the equilibrium Fermi distribution function, $\tau$ is the scattering time, $v_j$ is the velocity in the *j*-th direction, and $\mathbf{B} = B\vec{e}_z$ is a magnetic field which we take to be in the *z*-direction:

$$L_{xx}^{EE} = e^2 \int \frac{d^3k}{(2\pi)^3} \left(1 + \frac{e}{\hbar c}\mathbf{B}\cdot\mathbf{\Omega}\right)^{-1} \left(-\frac{\partial f_0}{\partial E}\right) \tau v_x^2 \frac{1}{1 + \frac{\omega_c^2 \tau^2}{\left(1 + \frac{e}{\hbar c}\mathbf{B}\cdot\mathbf{\Omega}\right)^2}} \tag{11}$$

$$L_{xy}^{EE} = e^2 \int \frac{d^3k}{(2\pi)^3} \left(1 + \frac{e}{\hbar c}\mathbf{B}\cdot\mathbf{\Omega}\right)^{-1} \left(-\frac{\partial f_0}{\partial E}\right) \tau v_y^2 \frac{\omega_c \tau}{\left(1 + \frac{e}{\hbar c}\mathbf{B}\cdot\mathbf{\Omega}\right)^2 + \omega_c^2 \tau^2} \tag{12}$$

$$L_{xx}^{ET} = -e \int \frac{d^3k}{(2\pi)^3} \left(1 + \frac{e}{\hbar c}\mathbf{B}\cdot\mathbf{\Omega}\right)^{-1} \frac{(E-\mu)}{T} \left(-\frac{\partial f_0}{\partial E}\right) \tau v_x^2 \frac{1}{1 + \frac{\omega_c^2 \tau^2}{\left(1 + \frac{e}{\hbar c}\mathbf{B}\cdot\mathbf{\Omega}\right)^2}} \tag{13}$$

$$L_{xy}^{ET} = -e \int \frac{d^3k}{(2\pi)^3} \left(1 + \frac{e}{\hbar c}\mathbf{B}\cdot\mathbf{\Omega}\right)^{-1} \frac{(E-\mu)}{T} \left(-\frac{\partial f_0}{\partial E}\right) \tau v_y^2 \frac{\omega_c \tau}{\left(1 + \frac{e}{\hbar c}\mathbf{B}\cdot\mathbf{\Omega}\right)^2 + \omega_c^2 \tau^2} \tag{14}.$$

These integrals can be evaluated to obtain Eq. 8. The chemical potential can be obtained by solving self-consistently as a function of temperature and assuming a constant density.

**Acknowledgments**

**Funding:** SJW is supported by the U. S. National Science Foundation Graduate Research Fellowship Program under grant number DGE-0822215. TMM, AP, and JPH are supported by the Center for Emergent Materials, an NSF MRSEC, under grant number DMR-1420451. JPH is also supported by the U. S. Army Research Office MURI under grant number W911NF-14-0016. NT is supported by funding from NSF-DMR-1309461. **Author contributions:** SJW and JPH designed and performed all experiments and analyzed the data. TMM and NT designed and implemented the theoretical numerical and analytical models. S-CW and YS completed DFT calculations. CS provided the samples. CF oversaw DFT calculations and sample preparation. AP assisted with measurements of the Nernst effect. SJW, TMM, JPH, and NT wrote the manuscript with input from all authors. **Competing interests:** The authors declare that they have no competing interests.

**Main Text Tables and Figures**

| | Period (T) | Fermi Surface Area ($10^{17}$ m$^{-2}$) | $k_F$ ($10^8$ m$^{-1}$) |
|---|---|---|---|
| Shubnikov-de Haas In Nernst H//c | 28.1 ± 0.6 | 2.70 | 2.93 |
| | 13.7 ± 0.8 | 1.31 | 2.03 |
| | 5.91 ± 0.7 | 5.64 | 1.34 |
| de Haas-van Alphen in magnetization H//c | 32.2 ± 0.5 | 3.10 | 3.13 |
| | 13.6 ± 0.4 | 1.30 | 2.03 |
| | 6.14 ± 0.05 | 5.86 | 1.36 |
| de Haas-van Alphen in magnetization H⊥c, 1 | 4.35 ± 0.4 | 4.15 | 1.15 |
| | 0.75 ± 0.08 | 0.70 | 0.47 |
| de Haas-van Alphen in magnetization H⊥c, 2 | 4.34 ± 0.4 | 4.14 | 1.15 |
| | 0.71 ± 0.15 | 0.68 | 0.46 |

**Table 1: Shubnikov-de Haas oscillations and de Haas-van Alphen quantum oscillations.** The Period, Fermi surface area, and Fermi momentum (assuming that the cross-sectional area is a circle) for the observed Shubnikov–de Haas oscillations in the Nernst effect (*H*//c-axis) and for the de Haas–van Alphen oscillations observed in the magnetization with *H*//c-axis and two orthogonal directions of *H* ⊥ c-axis are summarized for all three periods found in both measurements, which were taken on two separate samples of NbP. Results are similar between both data sets.

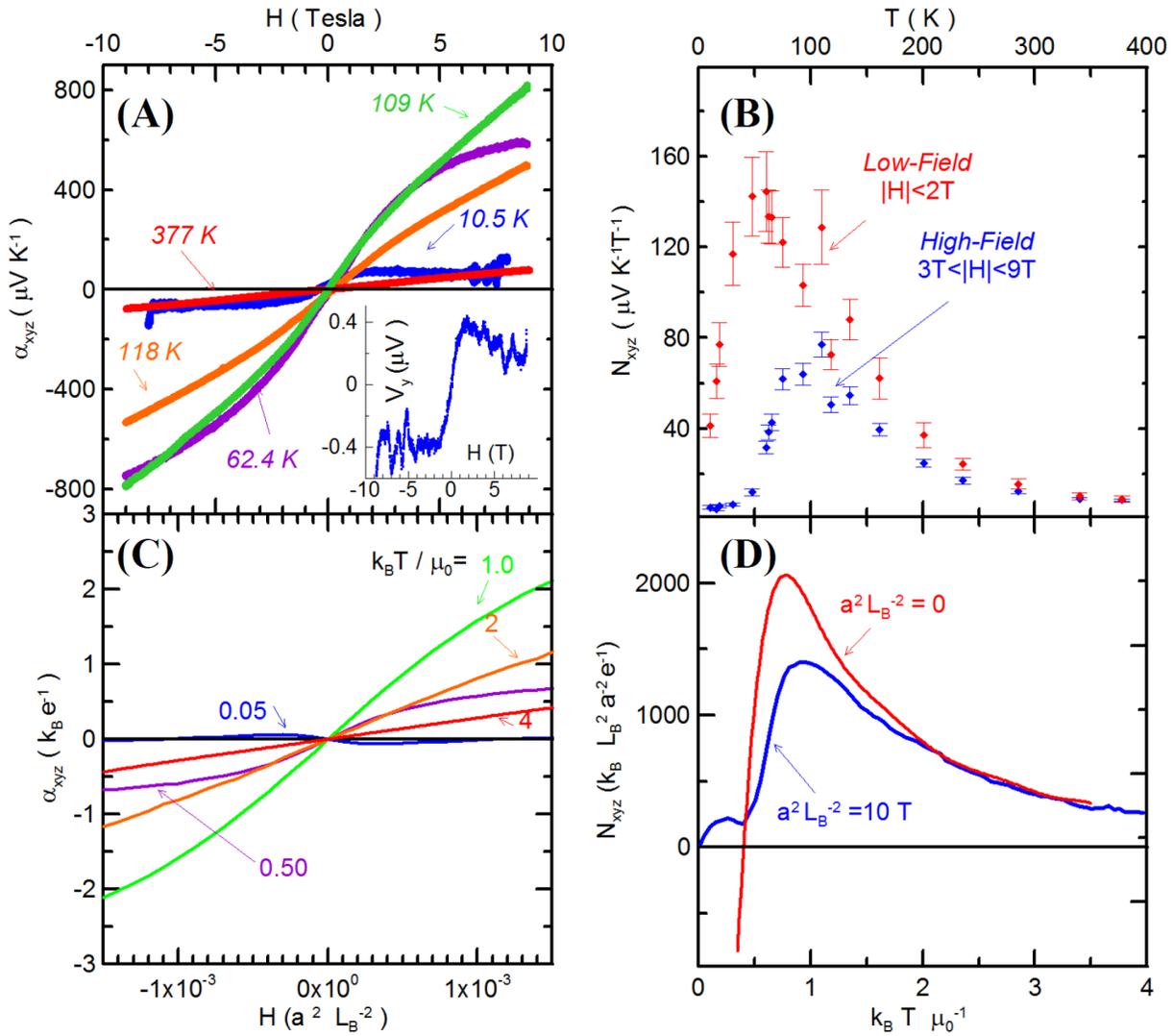

**Figure 1: Magnetic field dependence of Nernst thermopower, $\alpha_{xyz}$, and temperature dependence of Nernst coefficient, $N_{xyz}$.** (A) Data for the Nernst thermopower plotted as a function of applied magnetic field at the various temperatures indicated. The insert shows the magnetic field dependence of the Nernst voltage measured at 4.92 K in a temperature gradient of 2.17 mK. Shubnikov–de Haas oscillations are plainly visible that correspond to the periods measured in the magnetization. (B) Nernst coefficient plotted as a function of temperature, with the low-field Nernst effect in red and high-field Nernst effect in blue. The low-field curve peaks near 50 K; the high-field curve peaks near 90 K, which is the temperature at which the chemical potential touches the Weyl nodes. Error bars represent a 95% confidence interval on the standard deviation of the systematic error, excluding geometric error on the sample mount itself. (C) Calculated magnetic-field dependence of the Nernst thermopower at a few values of temperature. The values are given in natural units: the magnetic field as the product of the in-plane lattice

Page 14 of 38

constant $a$ and the magnetic length $L_B \equiv \sqrt{\hbar c / eB}$ ; the thermopower as $k_B/e$, and the temperature in units of $\mu_0/k_B$. (**D**) Calculated temperature dependence of the Nernst coefficient at zero field (red) and at 10 T (blue). As in (C), the theoretical values are given in natural units as indicated. Frames (A) and (C) and (B) and (D) are aligned to each other since $a$ and $\mu_0$ are measured independently. In both, a low-field and a high-field regime are visible with the different slopes representing the Nernst coefficients for each regime.

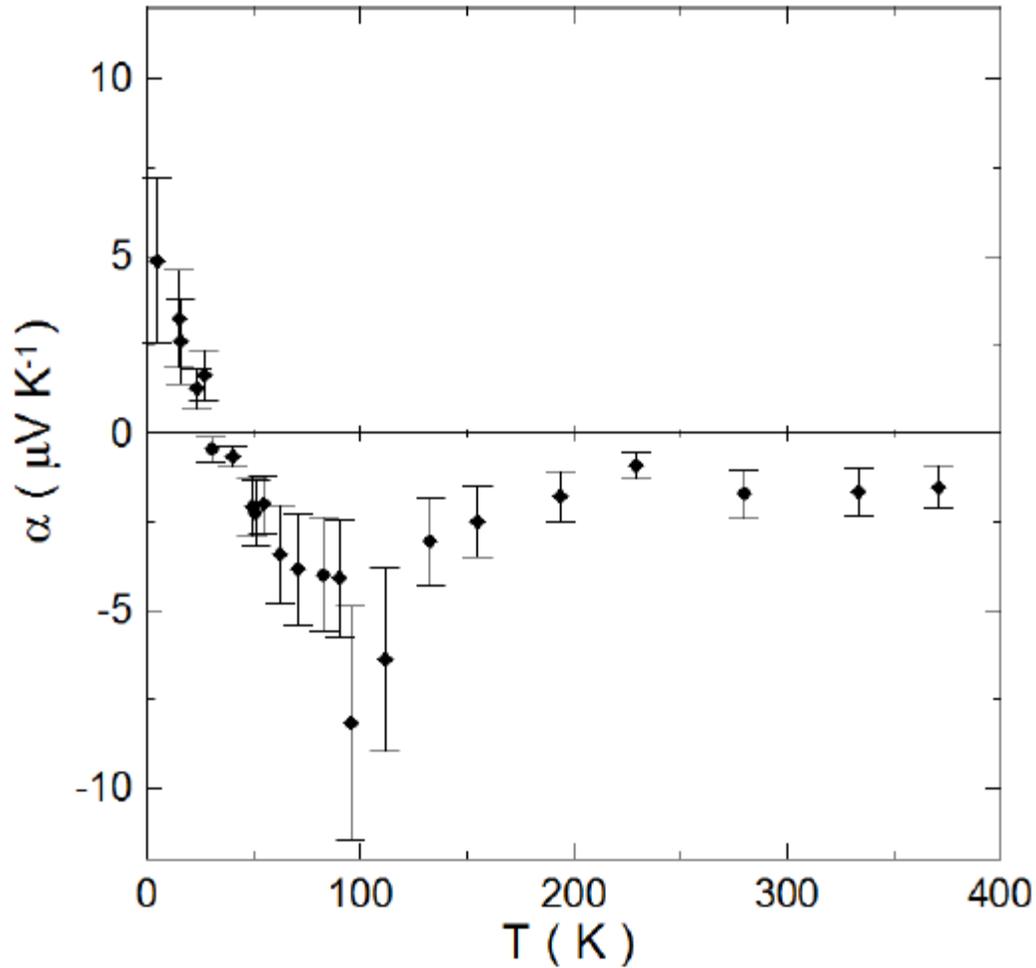

**Figure 2: Temperature dependence of thermopower in NbP.** Conventional thermopower is at maximum magnitude approximately 8 µV/K, over two orders of magnitude smaller than the maximum Nernst thermopower. A broad, negative minimum is present from 60 to 100 K. No magneto-thermopower is observed when a 9 T magnetic field is applied parallel to z, the <001> crystal axis. Error bars represent a 95% confidence interval on the standard deviation of the systematic error.

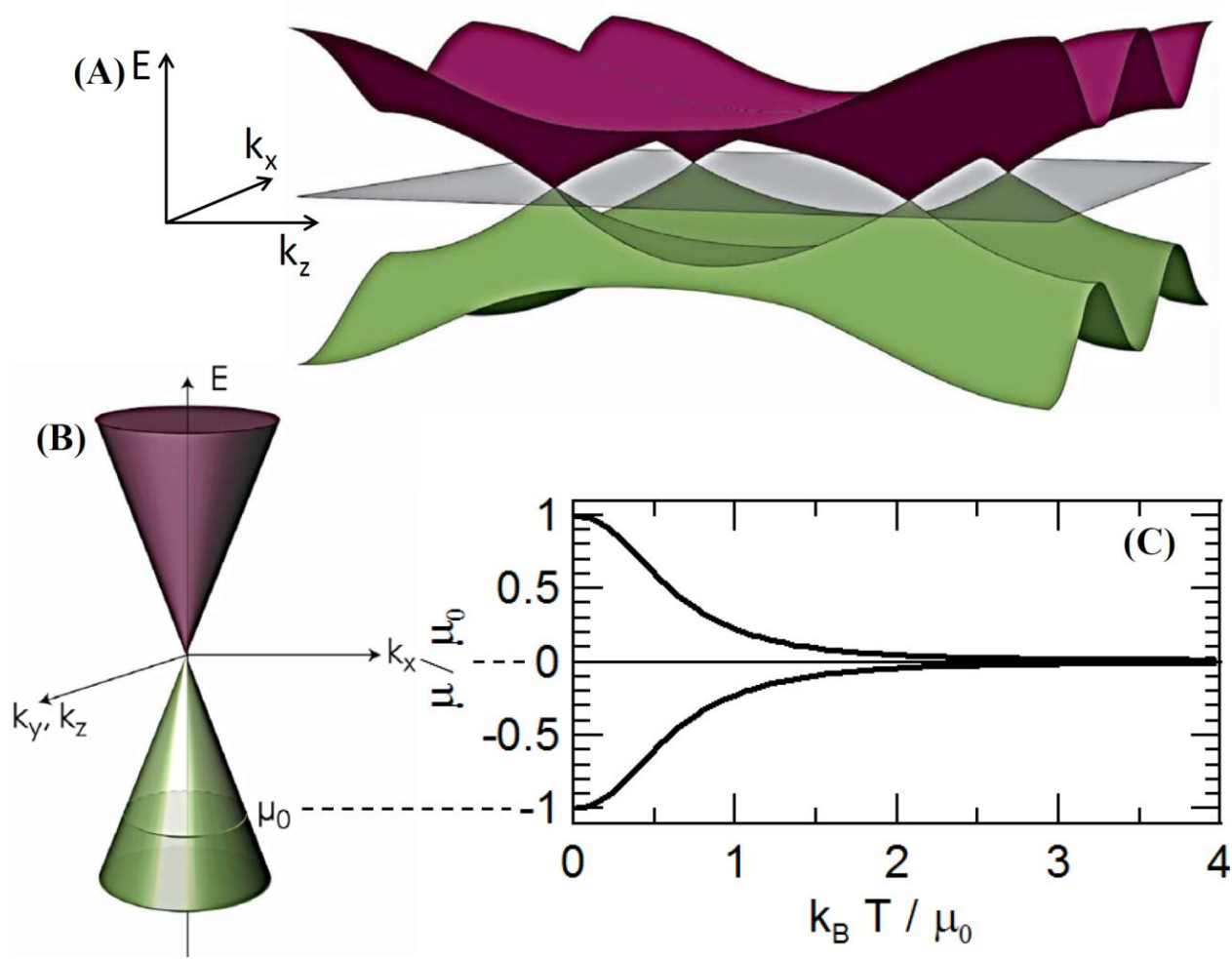

**Figure 3: Band structure and motion of the electrochemical potential with temperature.** (**A**) The two-dimensional electronic dispersion relation derived from Eq. 3 is used in modeling the transport properties of NbP. (**B**) The two-dimensional dispersion around a single Dirac cone is shown here. At low temperatures, the electrochemical potential lies in the valence band at an energy $\mu_0$ below the Dirac point, which is chosen as the origin of the energy scale. (**C**) The calculated temperature-dependence of the electrochemical potential $\mu(T)$ demonstrates the chemical potential moving towards the energy of the Weyl node with increasing temperature. The energy scale is set to 1 at the value of $\mu_0$, and the temperature scale is set to unity at the value of $\mu_0/k_B$.

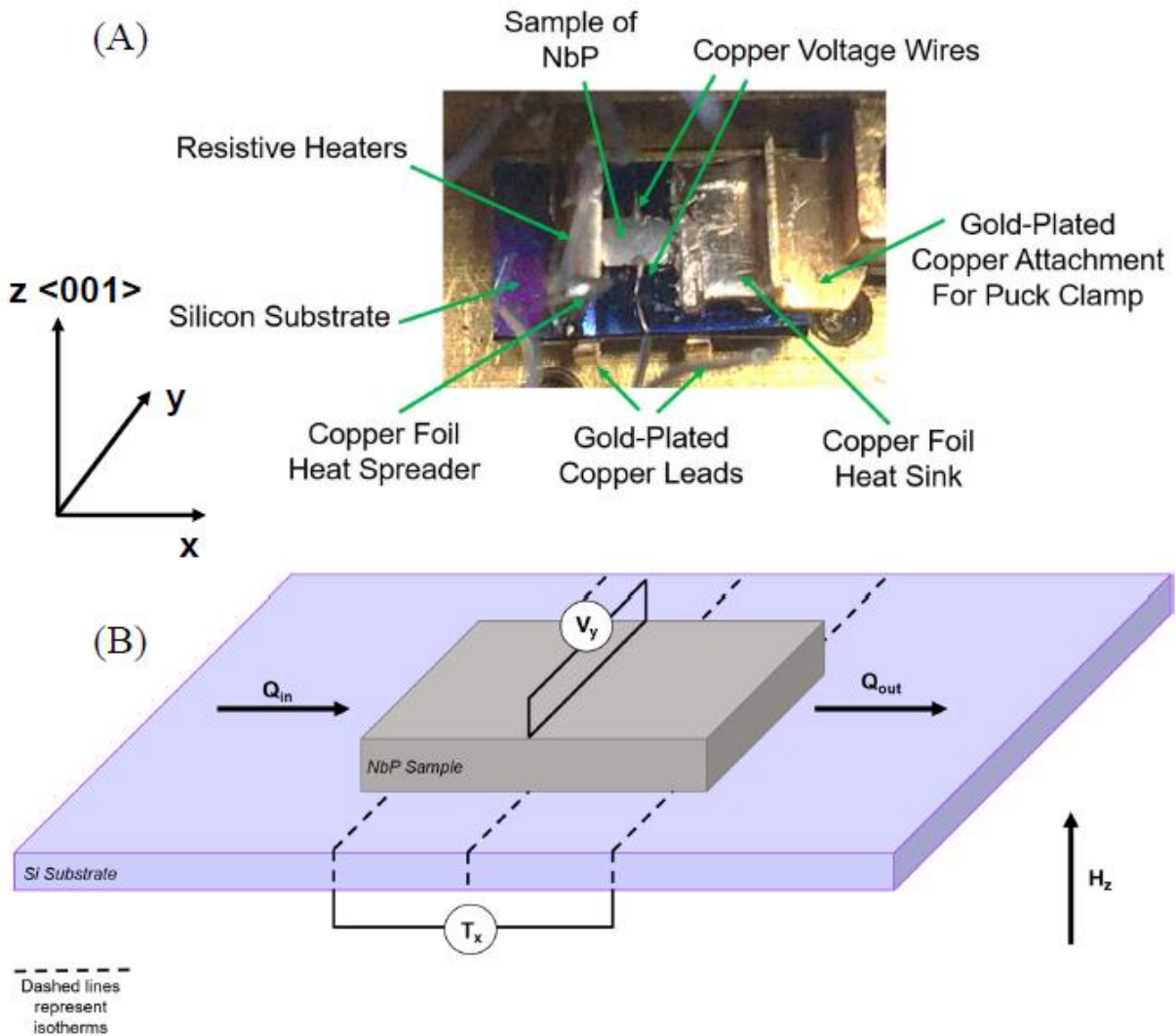

**Figure 4: Isothermal mount for Nernst and Seebeck Measurements in an externally applied magnetic field**. (**A**) A top view of the sample mounted on a Quantum-Design TTO puck shows the use of a silicon substrate, resistive heaters, copper voltage wires, and Cernox$^{TM}$ shoes from Quantum Design for temperature measurements. (**B**) The heat flow schematic of an isothermal sample mount displays the isotherms as dashed lines parallel to the measured electric field. The temperature gradient is applied parallel to *x*, voltage is measured parallel to *y* along isotherms, and external magnetic field is applied parallel to *z* which is aligned with the <001> crystal axis.

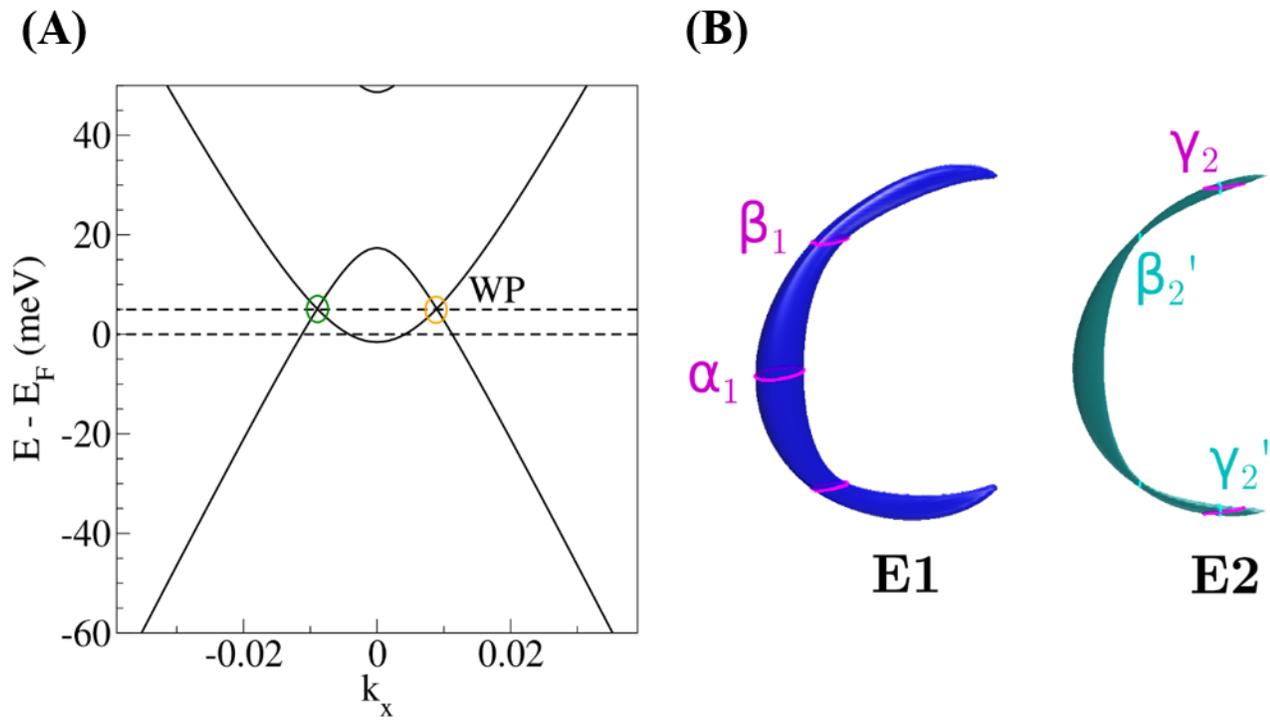

**Figure 5: Fermi surface from DFT.** (**A**) Energy dispersions are shown crossing the two Weyl points along $k_x$ direction. The chirality of Weyl points are highlighted by the yellow and green circles. (**B**) Fermi surfaces of two electron pockets, labeled as E1 and E2, are shown with their corresponding cross sections.

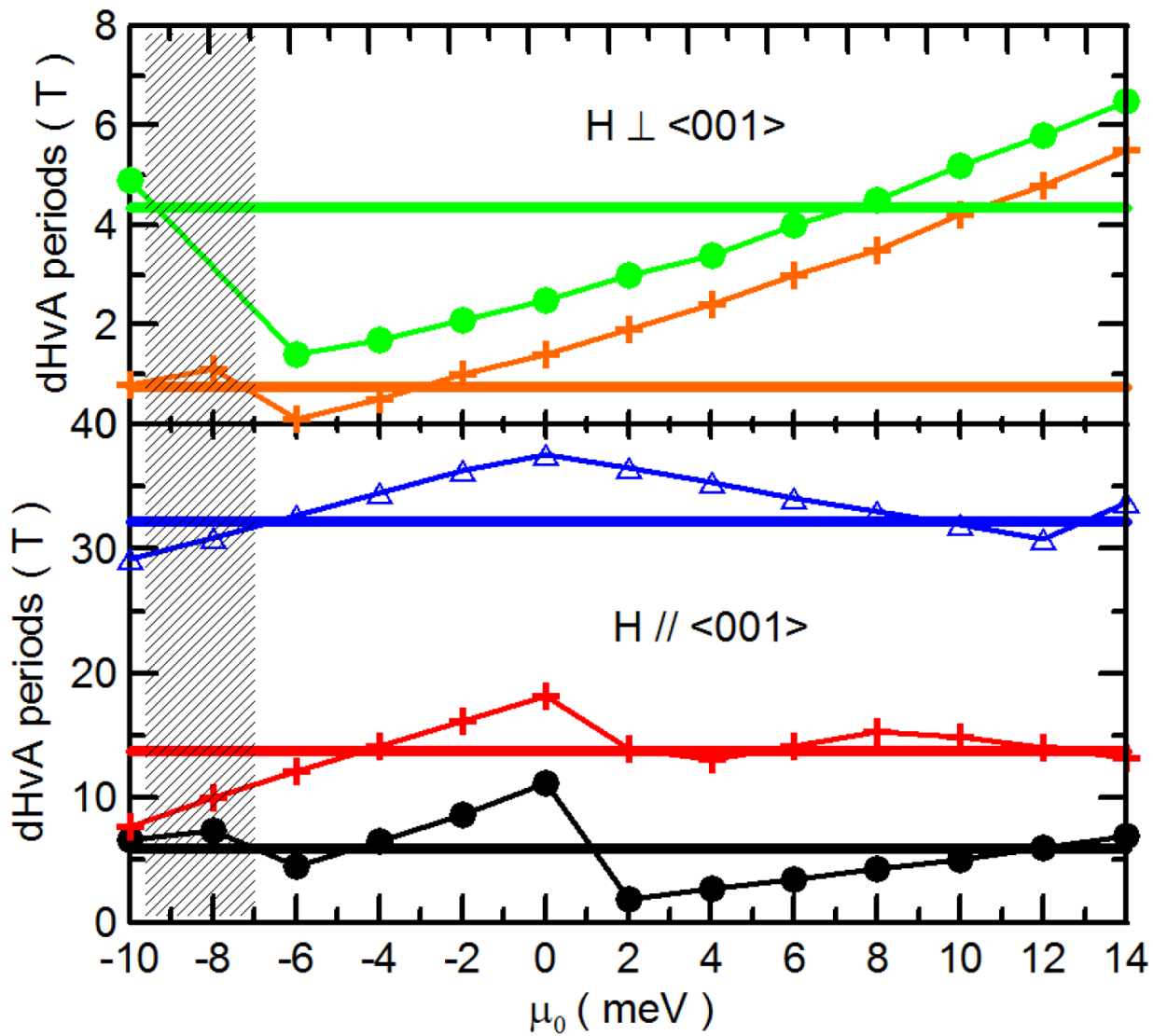

**Figure 6: Electrochemical potential at 0 K from dHvA periods and DFT calculations.**
Calculated (data points) and experimental (horizontal lines) values of the periods of de Haas-van Alphen oscillations are shown as a function of the electrochemical potential $\mu_0$ in the 0 K limit, as measured from the Dirac point of W1. The values of $\mu_0 = -8.2\pm0.7$ meV are represented by the hatched region, and the best overall fit falls in that box.

**Supplementary Materials**

*Adiabatic versus isothermal thermomagnetic measurements*

The differences between adiabatic and isothermal Nernst and magneto-Seebeck effects are well known in thermoelectric transport theory and discussed at length in Ref. (*29*). In order to ascertain experimentally that the large magnetic field dependence of the adiabatic Seebeck coefficient measured in Ref. (*27*) was indeed due to the adiabatic sample mount, we measured adiabatic transport properties of a second sample of single-crystal NbP of the same batch as the sample used in the main text, with dimensions of 1.78 mm x 1.67 mm x 0.25 mm. Measurements were taken in a home-built, continuous-flow cryostat. This instrument was cooled using liquid nitrogen with a maximum externally applied magnetic field of approximately 1.03 T. The magnetic field was applied parallel to the thin direction (labeled *z* in the experimental data figures) corresponding to the <001> crystal axis. The heat flux was applied along the *x*-axis, while the measurements of electric fields were either along the *x*-axis for the transverse Seebeck coefficient $\alpha_{xxz}$ or along the *y*-axis for the Nernst thermopower $\alpha_{xyz}$. The (*x,y,z*) coordinates are orthonormal, but the orientations of *x* and *y* in the <001> plane are not identified. In contrast to the sample used in the main text, which was attached to a silicon base, this sample was attached to a glass base plate with rubber cement. Silicon has a higher thermal conductivity than NbP, while glass has a much lower one. Consequently, the mount used here no longer measures the isothermal magneto-Seebeck and Nernst thermopowers, but thermomagnetic properties closer to the adiabatic ones (see Materials and Methods in the main text). Two copper-constantan thermocouples were attached to the sample using silver epoxy parallel to the x-axis of the sample, and two copper voltage wires were attached using silver epoxy opposite the thermocouples. A 120 Ω resistive heater was attached to a copper foil heat spreader using silver epoxy, which was then attached to the base plate and sample using silver epoxy as well. A copper foil heat sink was placed opposite the heat source, again using silver epoxy to attach it to the sample and base plate. Controls software was programmed using LabVIEW, and Keithley nanovoltmeters and precision current sources were used as instrumentation.

The zero-field properties in both the data shown here and in the main text reasonably agree, considering that two different instruments and samples were used. In contrast, they depart greatly when a magnetic field is applied parallel to the <001> crystal axis (*xxz* geometry) – the adiabatic data shows a strong magnetic field dependence below 150 K where the isothermal data shows no magnetic field dependence. The adiabatic data are shown in Fig. S1A and are similar to those of Stockert et al. (*27*) measuring similar single crystals of NbP adiabatically. When magneto-thermopower is measured adiabatically in a material with a Nernst thermopower significantly larger than conventional thermopower, the pure magneto-thermopower becomes unrecoverable adiabatically because the correction would be much larger than the signal itself. Thus, we conclude that all data sets are consistent, that there is no magneto-thermopower in the isothermal *xxz* geometry, and that the magnetic-field dependence of the adiabatic Seebeck coefficient observed in Ref. (*27*) is due to the development of a temperature gradient along *y*.

This difference between adiabatic and isothermal thermomagnetic transport measurements appears in the Nernst effect as well, although the problem is much less critical when the Nernst thermopower is significantly larger than the conventional thermopower, as is the case with NbP. Tsidil'kovskii developed approximate formulae to quantify the proportionality between isothermal and adiabatic electric fields measured in the presence of mutually orthogonally applied temperature gradients and applied magnetic fields, as shown here (*41*):

$$E_{y,adiabatic} = E_{y,isothermal}\left(1+\frac{7}{4}\frac{e\alpha_{xyz}}{k_B}\frac{\kappa_E}{\kappa_E+\kappa_L}\right), \mu_m B \ll 1$$
$$E_{y,adiabatic} = E_{y,isothermal}\left(1+\frac{45\pi}{64}\frac{e\alpha_{xyz}}{k_B}\frac{\kappa_E}{\kappa_L}\right), \mu_m B \gg 1$$
(S1)

where $E_{y,adiabatic}$ is the measured adiabatic electric field, $E_{y,isothermal}$ is the measured isothermal electric field, $\alpha_{xyz}$ is the Nernst thermopower, $\kappa_e$ is the electronic thermal conductivity, $\kappa_L$ is the lattice thermal conductivity, and *B* is the externally applied magnetic field. Eq. S1 is only applicable when the corrections are small (<15%). In the presence of an externally applied magnetic field, the adiabatic electric field deviates farther from the isothermal electric field as electronic thermal conductivity and Nernst thermopower increase. The pure Nernst thermopower or Nernst coefficient, as calculated by the Boltzmann transport theory, utilizes only the isothermal electric field. This is the experiment that was executed using the isothermal sample mount, which is described in the main text and shown in Fig. 4. We show the adiabatic Nernst thermopower in Fig. S1B. When we apply the correction of Eq. S1, we find correction factors of the order of 10% at 100 K, indicating the correspondence between Fig. S1B and Fig. 1 in the main text is quite within the experimental error. In contrast, the deviation of the adiabatic Seebeck thermopower in magnetic field from the isothermal Seebeck thermopower is too large to be corrected by the use of Eq. S1. The conversion between adiabatic and isothermal transport coefficients can rigorously be made via the Heurlinger relations *(29)*, but they necessitate a measurement of the Righi-Leduc coefficient, which was not possible due to the small dimensions of the sample.

*Sample characterization: electrical conductivity*

Fig. S2 shows the temperature dependence of the sample's electrical conductivity as well as the transverse magnetic field dependence and the Hall resistivity (magnetic field applied parallel to <001> in both cases), which were taken using the adiabatic sample mount technique described previously. We report an increase in resistivity with increasing temperature. We also observe an increase in transverse magneto-resistivity with magnetic field, with magnetic field dependence becoming stronger as temperature decreases towards 100 K. From Hall resistivity, we can infer a peak in the Hall coefficient near 160 K. Magnetoresistance and Hall effect measurements are given in Fig. S2 for comparison with earlier work.

Although samples used in this study and in that of Shekhar et al. (*17*) are similar, we see a discrepancy in electrical resistivity between the two samples, indicating that our samples are on the order of ten times less resistive. A large contribution to this must come from the geometric uncertainty in the measurements. Due to the small size of the sample, a geometric error on the order of hundredths of a millimeter results in a noticeable offset in resistivity data. Measuring the distance separating the voltage wires on the sample was not possible with this high level of precision, as the wires were not placed directly on the edges of the sample, and getting this precision with micrometers actually could result in breaking the wires off of the sample.

*Sample characterization: thermal conductivity*

Fig. S3 reports the temperature dependence of the sample's thermal conductivity in NbP, with data taken using the adiabatic sample mounting technique described previously. The data show a decrease in thermal conductivity with increasing temperature. This data compares quite well to that of Stockert et al. (*27*), as both studies use similar single crystals of NbP. As mentioned in the previous section, there is considerable geometric uncertainty in both this and resistivity measurements due to the small size of the sample; however, the electrical and thermal conductivity reported in this supplement use the same geometric measurements and are internally consistent.

*Sample characterization: specific heat*

Specific heat data was taken on a single-crystal sample of NbP using the Heat Capacity Option on a Quantum Design PPMS. Data was taken between 2 K and 400 K, and a $2\tau$ fitting method was used. Fig. S4 shows the specific heat versus temperature, and the data are consistent with the calculation in the literature (*42*). In frame S4A, the full curve is shown, fitting a Debye model with a Dulong-Petit limit of 49.9±0.2 J/(NbP formula unit)-K and a Debye temperature $\Theta=525\pm2$ K. Fig. S4B shows the separation of the specific heat into a linear term and a $T^3$ term following the equation:

$$C(T) = \gamma T + AT^3 \qquad (S2)$$

where $C$ is the sample specific heat, $T$ is the temperature, $\gamma$ is the coefficient for the linear contribution, and $A$ is the coefficient for the cubic contribution. This plot was made by dividing the entire equation by the temperature, plotting $C/T = \gamma + AT^2$ as a function of $T^2$. The result gives $\gamma=1.5\ 10^{-4}$ J/mol-K$^2$ and A=2.7x10$^{-5}$ J/mol-K$^4$. Because the Dirac dispersion is linear and the DOS$(E) \propto E^2$, one expects the electronic specific heat of the electrons in the Dirac cone to follow a $T^3$ law, which is indistinguishable from the phonon contribution. Therefore, the $\gamma$-value reported corresponds presumably only to the charge carriers located in the trivial pockets in the zero-K limit since $\gamma$ can only be resolved below 10 K.

*Sample characterization: magnetization*

Magnetization data were measured using the Vibrating System Magnetometer (VSM) on a Quantum Design PPMS in an external magnetic field applied either $H\ //\ <001>$ or $H \perp <001>$, with data shown in Fig. S5. Two orientations with $H \perp <001>$ were studied normal to each other, but they were not aligned with the <100> or <010> axes; the results were isotropic in-plane. The sample was attached to the sample holder using GE Varnish. The magnetic moment of the sample was measured with the externally applied magnetic field sweeping both up and down from negative to positive field and vice-versa between maximum magnitudes of ±7 T.

The data indicate that NbP is a diamagnet with an isotropic susceptibility $\chi_d = -4.0\pm0.2$ $\mu_B$ Oe$^{-1}$ per NbP formula unit at 5K, with $\mu_B$ the Bohr magneton. Clear de Haas – van Alphen quantum oscillations are visible that were analyzed using a Fourier transform. The observed periods, Fermi surface areas (assuming a circular Fermi surface), and Fermi momenta are displayed in Table 1 of the main text. With the magnetic field applied parallel to the <001> crystal axis, three periods are present and correspond very well to those found from Shubnikov–de Haas oscillations in the Nernst measurements and from Shekhar et al.'s (*17*) data. The in-plane axes, perpendicular to the *c*-axis, show two periods present, and they are similar when comparing the two in-plane axes. The periods, Fermi surface areas, and Fermi momenta are reported in the main text in Table 1. The periods with $H\ //\ <001>$ also correspond quite well with those of the Shubnikov–de Haas oscillations in the Nernst effect. Stockert et al. (*27*) report more quantum oscillations in heat capacity, magnetization, thermal conductivity, and adiabatic magnetothermopower that, for $H \,//\, <001>$, all have periods of oscillations comparable to each other and comparable to those reported here, although those samples were different and presumably have slightly different values of $\mu_0$.

*Theoretical analysis of effects of multiple Weyl nodes and trivial pockets*

In this section, the effects of multiple Weyl nodes on the temperature dependence of the chemical potential and the Nernst thermopower are analyzed. A single Weyl node has a density of states given by:

$$g(E) = \frac{1}{\pi^2} \frac{E^2}{(\hbar v_F)^3} \tag{S3}$$

where $v_F$ is the Fermi velocity of the node and $E$ is the energy from the Weyl node. If multiple pairs of Weyl nodes exist, possibly at different energies and with different Fermi velocities, we can write:

$$g(E) = \sum_i \frac{N_i}{\pi^2} \frac{(E - E_i)^2}{(\hbar v_i)^3} \tag{S4}$$

where $N_i$ is the number of nodes in the $i$-th set, $E_i$ is the energy of the $i$-th set, and $v_i$ is the Fermi velocity of the $i$-th set of Weyl nodes. For a fixed density of states, the temperature dependence of the chemical potential can be determined. If all sets of Weyl nodes do not exist at the same energy, the chemical potential shifts to $E_m$, the energy for which Eq. S4 is minimum, rather than shifting to the node. For two sets of nodes, as is the case in NbP, this minimum occurs at:

$$E_m = \frac{\frac{N_1}{v_1^3} E_1 + \frac{N_2}{v_2^3} E_2}{\frac{N_1}{v_1^3} + \frac{N_2}{v_2^3}} \tag{S5}$$

Following the results of DFT calculations for NbP shown in Fig. 6, we take the velocities of the Weyl fermions to be isotropic, and we take the Fermi velocities of the two sets of Weyl nodes to be the same, $v_1 \approx v_2$. The $N_1 = 8$ nodes labeled W1 reside at $E_1 \approx -60$ meV. There are also $N_2 = 16$ nodes labeled W2 which lie at $E_2 \approx 6$ meV. We measure our energies in units of $E_1$ and find that $E_m \approx 0.63|E_1|$. By self-consistently solving for the chemical potential as a function of temperature, we find that on a temperature scale $T_m = E_m/k_B$, the chemical potential moves to the energy for which the density of states is minimum. For this choice of parameters, we show the density of states from Eq. S4 for these sets of Weyl nodes in Fig. S6A. We see the temperature dependence of the chemical potential in Fig. S6B shows µ($T$) shifting to the Weyl nodes at $T = T_m$.

We can calculate the components of the thermoelectric tensors $L_{ij}^{\alpha\beta}$ as we did in the main text. The Nernst thermopower for the same parameters as above is shown in Fig. S7. We have taken the magnetic field to be $B = 0.1\, cE_1^2 / e\hbar v_1^2$. We find that the peak in the Nernst thermopower $\alpha_{xyz}$ occurs where the density of states is minimized at $T_m$. We note that the value of the Nernst thermopower is about two times less than what we found in the case of a single set of Weyl nodes. This is due to the presence of a finite density of states, even at $E_m$.

Finally, we note that there are several differences between the simple model used in the main text and one that adds the effects of the trivial pockets. In NbP at 0 K, there are topologically trivial, massive bands. If these persisted at higher temperature, they would modify the density of states, and we expect that they would have contributed to thermoelectric transport in ways that do not correspond at all to what is observed experimentally. In the presence of trivial bands, the chemical potential still will move to the energy for which the density of states is minimized. We also expect that the peak of the Nernst effect at $T_m$ would be maintained as a general feature of Weyl semimetals, but the thermopower near room temperature would have picked up a metallic behavior. As stated in the main text, the fact that the simple theory explains the observations so well can be viewed as evidence for the fact that the trivial pockets moved out of the tens-of-meV range of the Dirac points due to the temperature dependence of the band structure. The band structure of all solids changes by an amount of the order of $10^{-2}$ eV between cryogenic and room temperatures: e.g. the band gap of PbTe increases almost linearly with temperature from 180 meV at 0 K to 350 meV at 400 K (*43*), whereas that of Ge decreases from 745 meV at 0 K to 610 meV at 400 K (*44*). There is no reason to expect NbP to behave differently, and the energy range of the changes observed is an order of magnitude larger than the energy range that matters in the present model.

Analytic functions were also developed per Eq. 10 for the zero-field Seebeck and the low-field Nernst coefficients. Results are shown in Fig. S8. They correspond very well to the full numerical calculations plotted in Fig. 1.

*Supplementary Materials Figures and Labels*

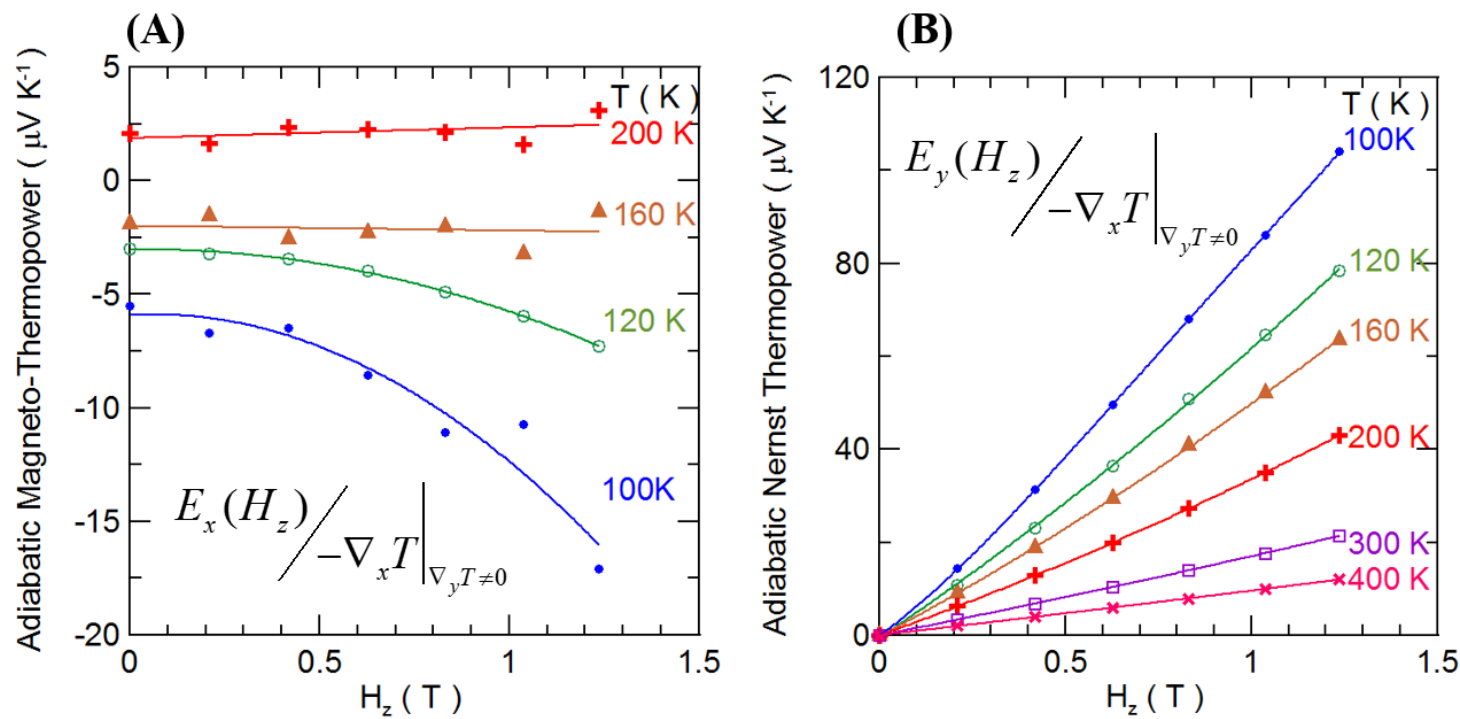

**Figure S1: Magnetic-field dependence of adiabatic Seebeck and Nernst thermopowers at select temperatures in NbP.** (**A**) The adiabatic magneto-thermopower shows a magnetic-field dependence at lower temperatures. This magnetic-field dependence is not measured in the isothermal mount described in the main text, implying that the magnetic-field dependence seen in the adiabatic mount is purely a result of the contribution of the Nernst thermopower to the adiabatic Seebeck effect. (**B**) The adiabatic Nernst thermopower shows an increase in signal with both increasing magnetic field and temperature up to 100 K (note that this instrument only covers the low-field regime described in the main text).

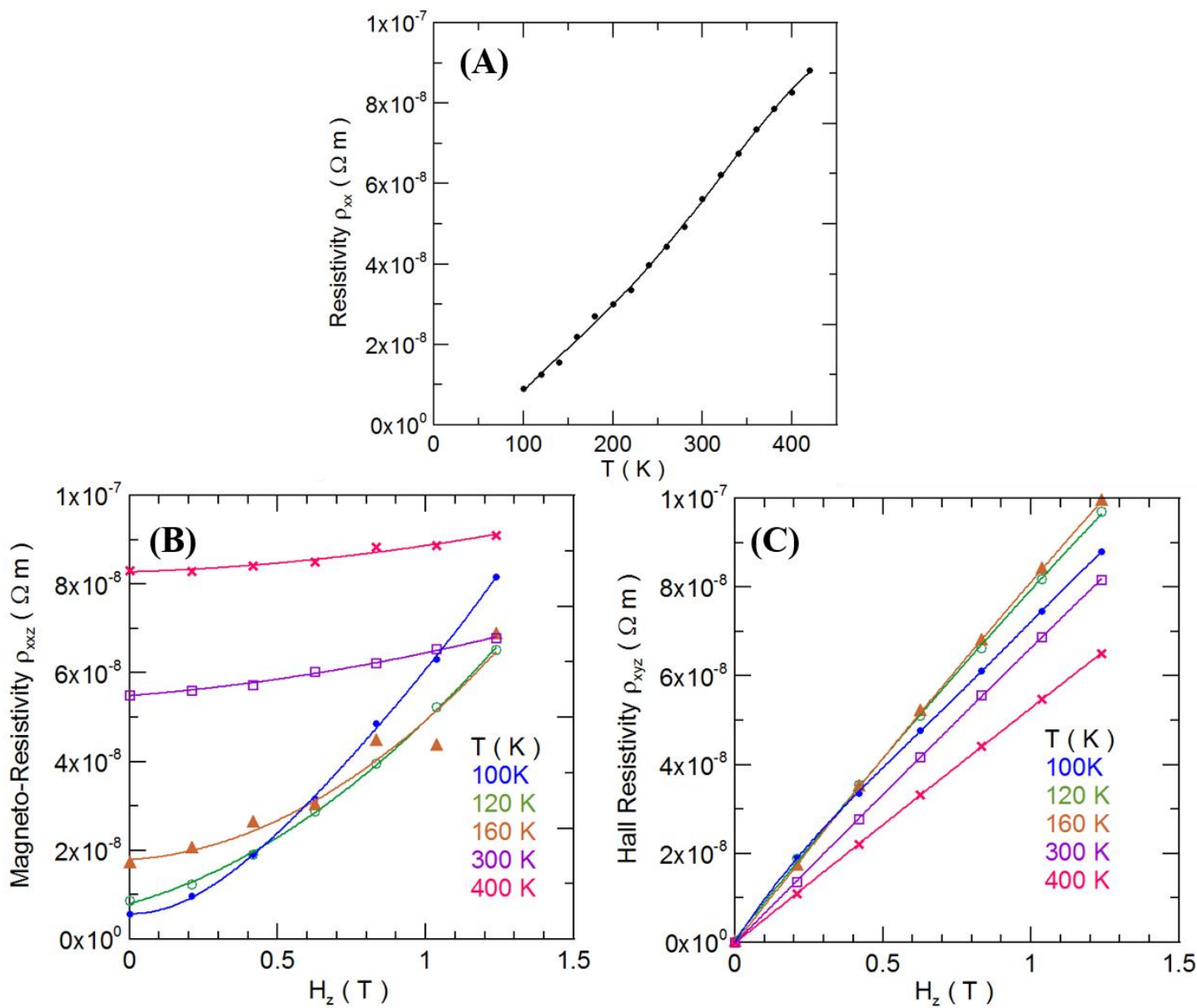

**Figure S2: Electrical transport in NbP.** (**A**) Resistivity demonstrates an increase with increasing temperature. (**B**) Transverse magneto-resistivity shows an increase in resistivity with both increasing temperature and magnetic field, where the field dependence becomes stronger at lower temperatures. (**C**) Hall resistivity demonstrates an increase in signal with increasing magnetic field and a non-monotonicity in temperature.

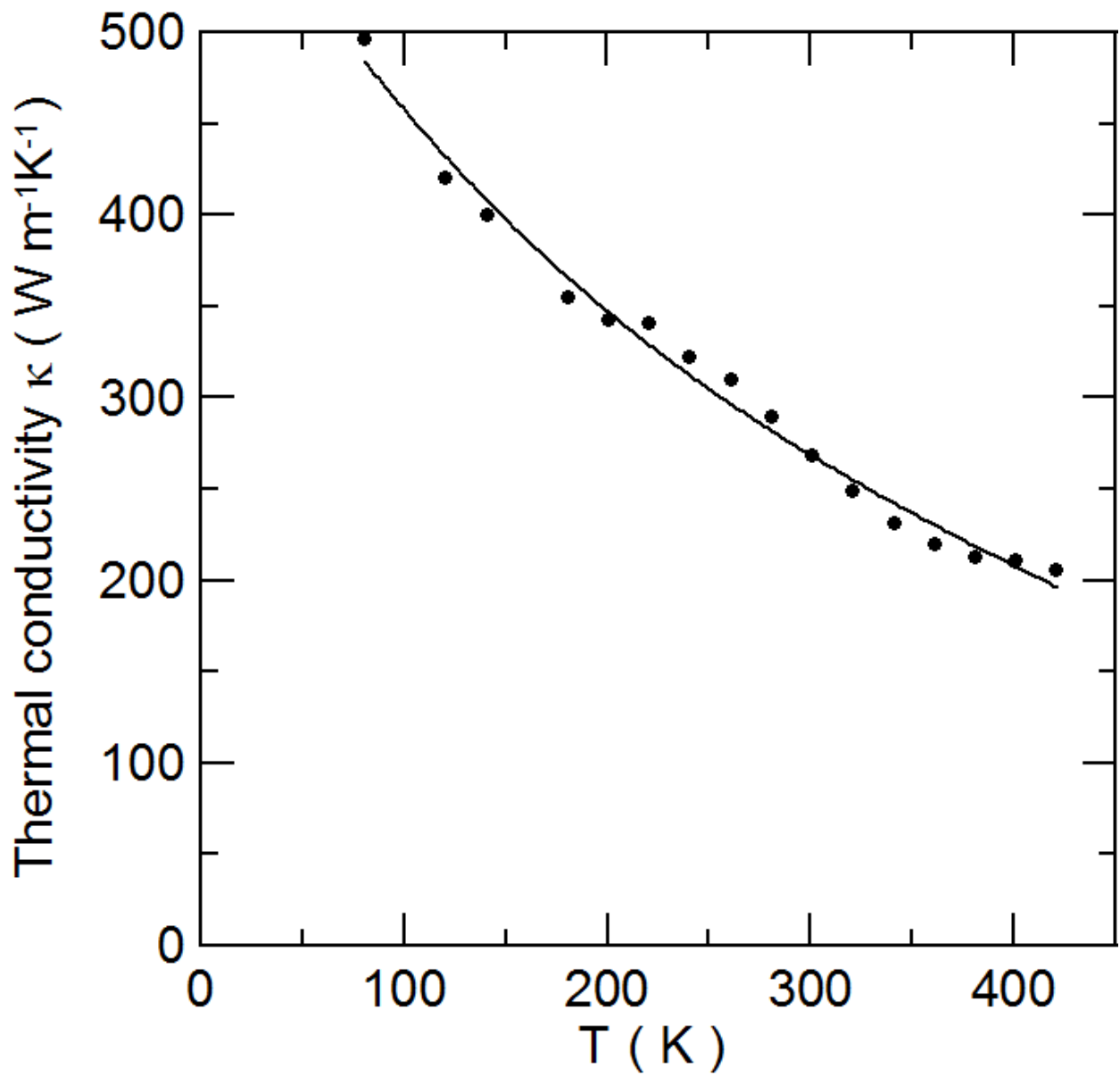

**Figure S3: Temperature dependence of thermal conductivity in NbP.** The thermal conductivity of NbP increases with decreasing temperature within the temperature range of this instrument.

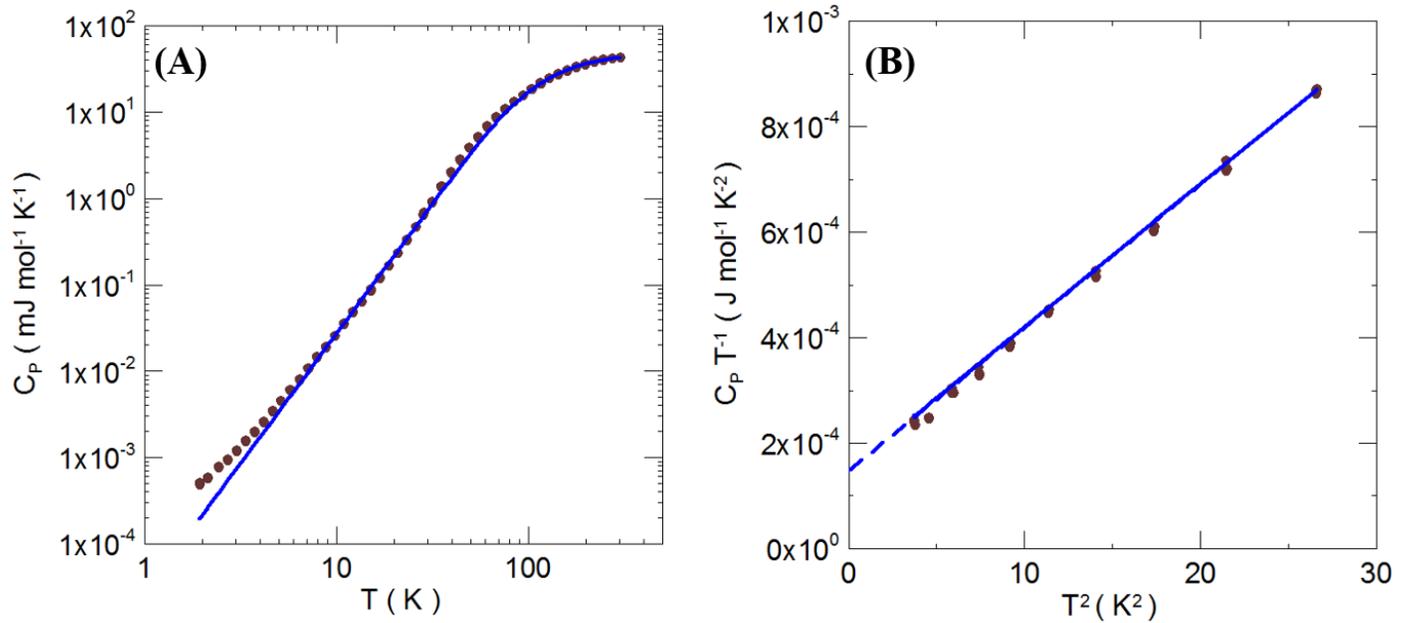

**Figure S4: Specific-heat analysis of NbP.** (**A**) The temperature dependence of specific heat (data are given as points) fits a Debye model (full blue line) with a Dulong-Petit limit of 49.9±0.2 J/mol K and a Debye temperature $\Theta=525\pm2$ K. (**B**) Low-temperature specific heat divided by temperature is plotted as a function of temperature squared to separate the specific heat into a linear term and a $T^3$ term, resulting in $\gamma=1.5\ 10^{-4}$ J/mol K$^2$, which is attributed only to the electrons in the trivial pockets of the Fermi surface. The electronic specific heat of the electrons in the Dirac cone is expected to follow a $T^3$ term, indistinguishable from the phonon term, due to the linear Dirac dispersion and the density of states following the square of energy.

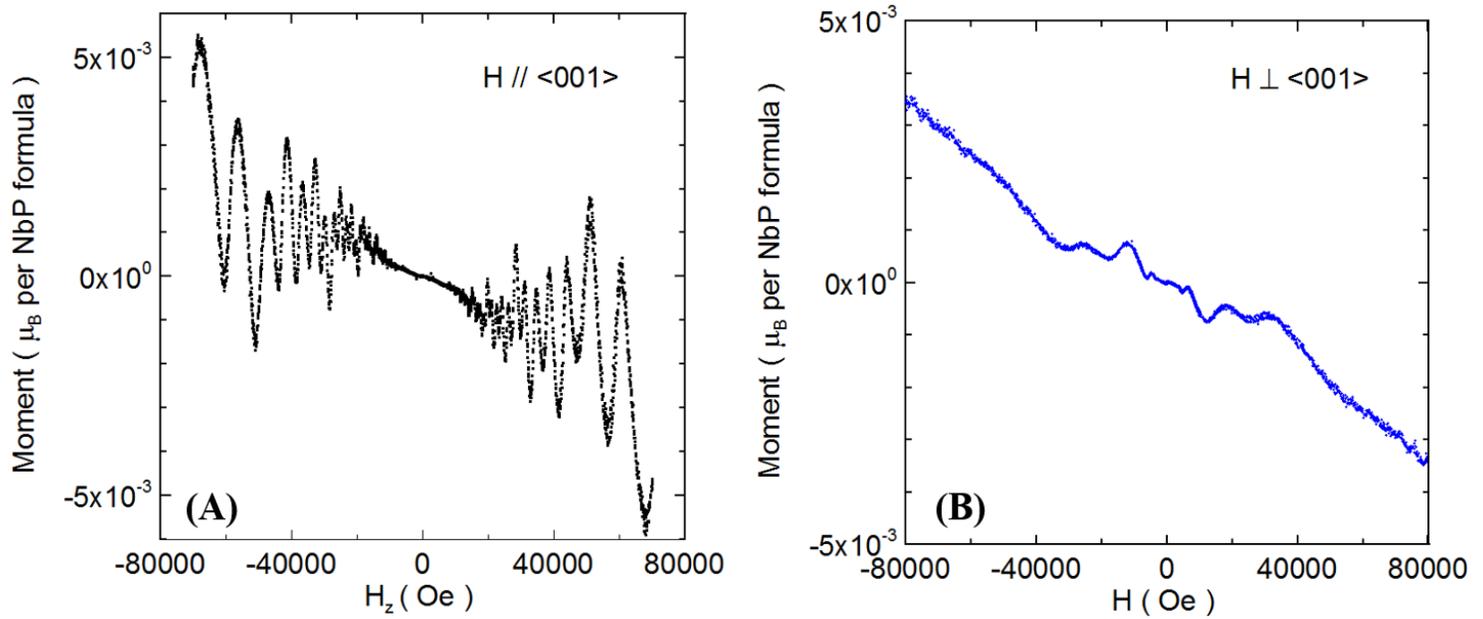

**Figure S5: Diamagnetic moment of NbP at 5K.** Magnetic moment data taken using VSM on a PPMS shows that NbP is a diamagnet in all directions, with data displayed showing the externally applied magnetic field parallel to the c-axis, which is the <001> crystal axis, in frame (**A**) and in the plane of <001> in frame (**B**). Two normal directions in the plane were examined with isotropic results. $\mu_B$ is the Bohr magneton. From this data, oscillation periods, Fermi surface area (assuming a circular Fermi surface), and Fermi momenta are derived and tabulated in Table 1 of the main text.

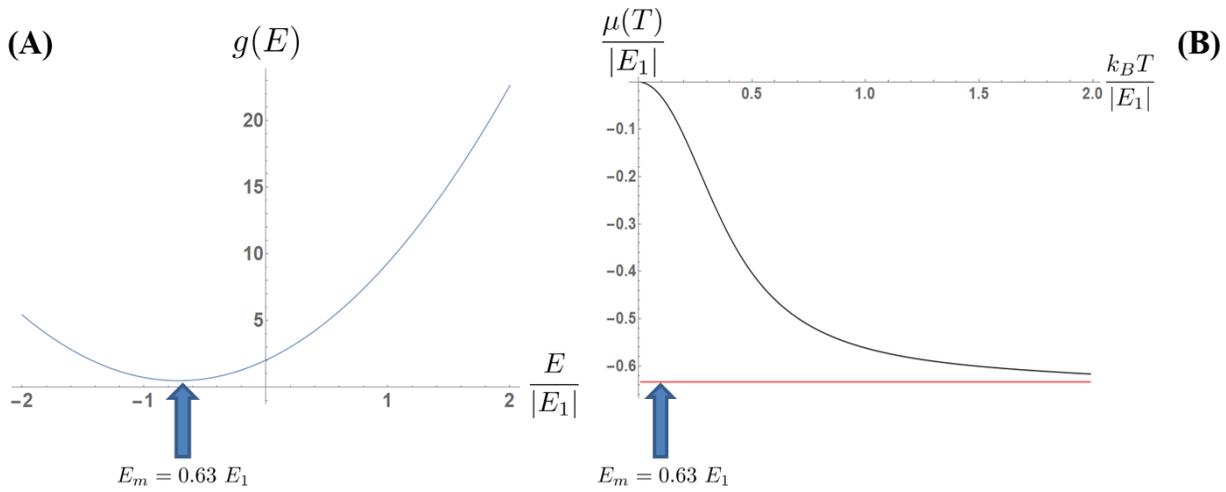

**Figure S6: Density of states to determine temperature dependence of the chemical potential.** (**A**) The density of states for $N_1 = 8$, $N_2 = 16$, and $E_2/E_1 = 0.1$ is shown. The minimum of density of states occurs at $E_m \approx 0.63 E_1$. (**B**) The temperature dependence of the chemical potential for parameters in (A) is displayed. We see that the chemical potential shifts to $E_m$ on a temperature scale $T_m = E_m/k_B$. $E_m$ is shown as a red line.

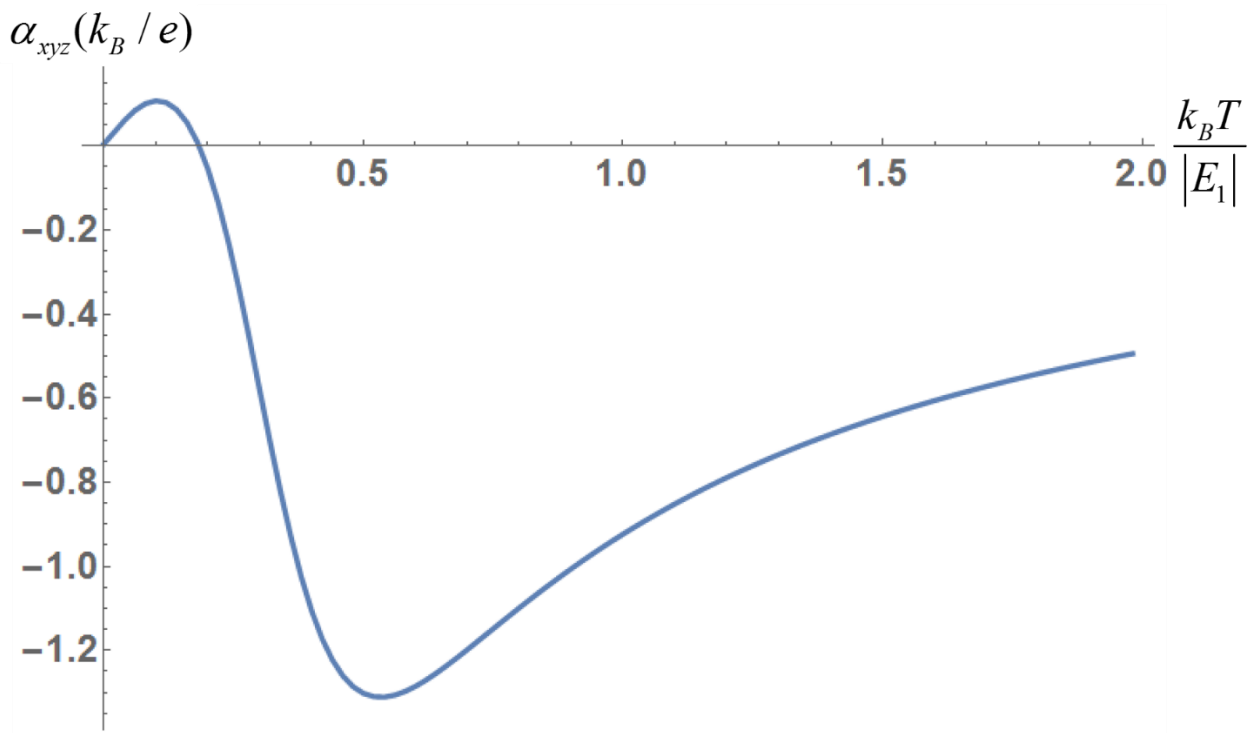

**Figure S7: Nernst thermopower as a function of temperature for $N_1 = 8$, $N_2 = 16$, and $E_2/E_1$ = 0.1.** The magnetic field is taken to be $B = 0.1 cE_1^2/e\hbar v_1^2$. The Nernst thermopower $\alpha_{xyz}$ has a maximum at the temperature scale $T_m = E_m/k_B$.

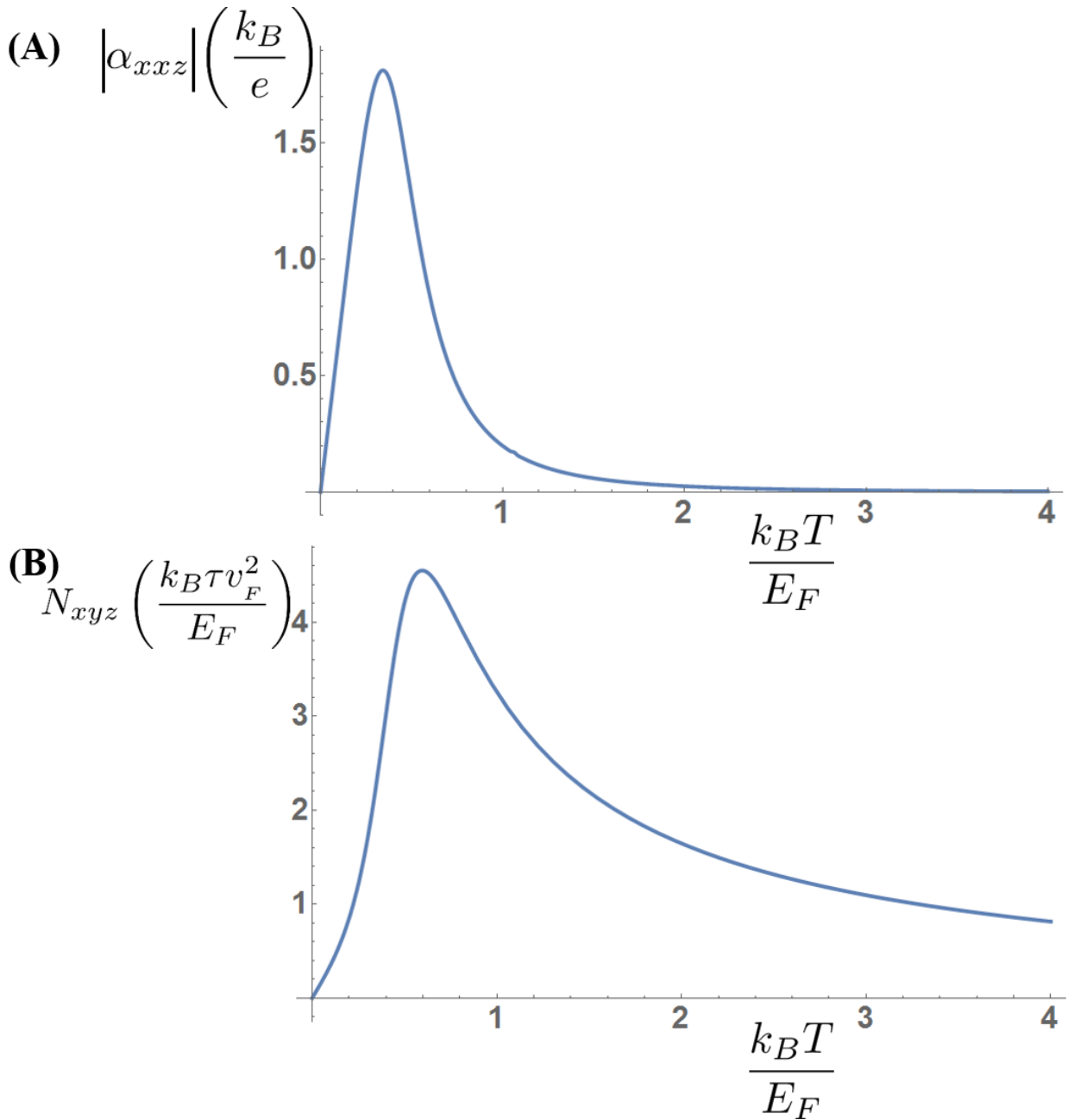

**Figure S8: Analytic zero-field absolute value of the Seebeck thermopower and low-field Nernst coefficient, both as a function of temperature.** (**A**) As calculated from Eq. 10, the absolute value of the conventional thermopower has decreased to a small fraction of the Nernst thermopower near and above $T_m$. (**B**) As calculated from Eq. 10, the experimental and numerical results of Fig. 1 show excellent agreement with the analytic results displayed here. From the value of the peak, a reasonable estimate for the value of the scattering time is determined to be $\tau = 10^{-13}$ s.